\documentclass[aps,showpacs, pra,amssymb, amsmath, twocolumn]{revtex4}
\usepackage{amsmath,latexsym,amssymb}
\usepackage{graphicx}
\usepackage{amsmath}
\usepackage{latexsym}
\usepackage{amssymb}
\usepackage{dcolumn}
\usepackage{bm}
\usepackage{placeins}
\usepackage{color}

\begin{document}

\title{Defect-free atomic array formation using Hungarian matching algorithm}
\author{Woojun Lee, Hyosub Kim, and Jaewook Ahn}
\email{jwahn@kaist.ac.kr}
\address{Department of Physics, KAIST, Daejeon 305-701, Korea}
\date{\today}

\begin{abstract}
Deterministic loading of single atoms onto arbitrary two-dimensional lattice points has recently been demonstrated, where by dynamically controlling the optical-dipole potential, atoms from a probabilistically loaded lattice were relocated to target lattice points to form a zero-entropy atomic lattice. In this atom rearrangement, how to pair atoms with the target sites is a combinatorial optimization problem: brute-force methods search all possible combinations so the process is slow, while heuristic methods are time-efficient but optimal solutions are not guaranteed. Here, we use the Hungarian matching algorithm as a fast and rigorous alternative to this problem of defect-free atomic lattice formation. Our approach utilizes an optimization cost function that restricts collision-free guiding paths so that atom loss due to collision is minimized during rearrangement. Experiments were performed with cold rubidium atoms that were trapped and guided with holographically controlled optical-dipole traps. The result of atom relocation from a partially filled 7-by-7 lattice to a 3-by-3 target lattice strongly agrees with the theoretical analysis: using the Hungarian algorithm minimizes the collisional and trespassing paths and results in improved performance, with over 50\% higher success probability than the heuristic shortest-move method.
\end{abstract}

\maketitle

\section{Introduction}

Neutral atom arrays in two or three dimensional space may play an important role in quantum information processing (QIP), because of their scalability to a massive number of qubits~\cite{RaussendorfPRL2001,IsenhowerPRL2010,WilkPRL2010,KaufmanNat2015,JauNP2016,SaffmanJPB2016}. Currently, arrays of several hundred atoms have been implemented with optically-addressable spacings of a few $\mu$m~\cite{NogrettePRX2014,PiotrowiczPRA2013}, and this number is expected to increase to a few thousand as laser power permits. These atoms are confined by an array of optical-dipole traps made through various methods including holographic devices~\cite{BergaminiJOSAB2004}, diffractive optical elements~\cite{KnoernschildAPL2010, Gillen-ChristandlPRA2011}, micro-lens arrays~\cite{DumkePRL2002}, and optical lattices~\cite{GreinerNat2002}. Ultimately, neutral-atom platforms for QIP may require (i) a significant number of atoms, (ii) a high-dimensional architecture, preferably with an arbitrary lattice geometry, (iii) single-atom loading per site, and (iv) the ability to be individually addressable. However, no existing method satisfies all these requirements. For example, optical lattices can provide a large number of atoms singly loaded per site through the Mott insulator transition~\cite{GreinerNat2002}, but they have rather limited geometries and often lack individual addressability; other methods have advantages of arbitrary configurations and site addressability but fail the single-atom loading condition due to the collisional blockade effect~\cite{SchlosserPRL2002}. 

In optical-dipole traps, the probability of single-atom trapping per site is about 50 percent. Both the filling factor and the configuration of the entire array are, in consequence, probabilistic. The probability of filling an entire array with $N$ atoms scales as $0.5^N$, which is extremely small for a large $N$. Significant efforts are being devoted to achieve a deterministic or near-deterministic single-atom loading; one approach uses an array of bottle-shaped blue-detuned optical well potentials~\cite{XuOL2010}, and the others include light-assisted, controlled inelastic collision~\cite{GrunzweigNPhys2010,LesterPRL2015,FungNJP2015}. The loading probability of defect-free arrays however still remains distant from one, especially when we consider a large number of atoms. 

\begin{figure}[t]
\centering
\includegraphics[width=0.4\textwidth]{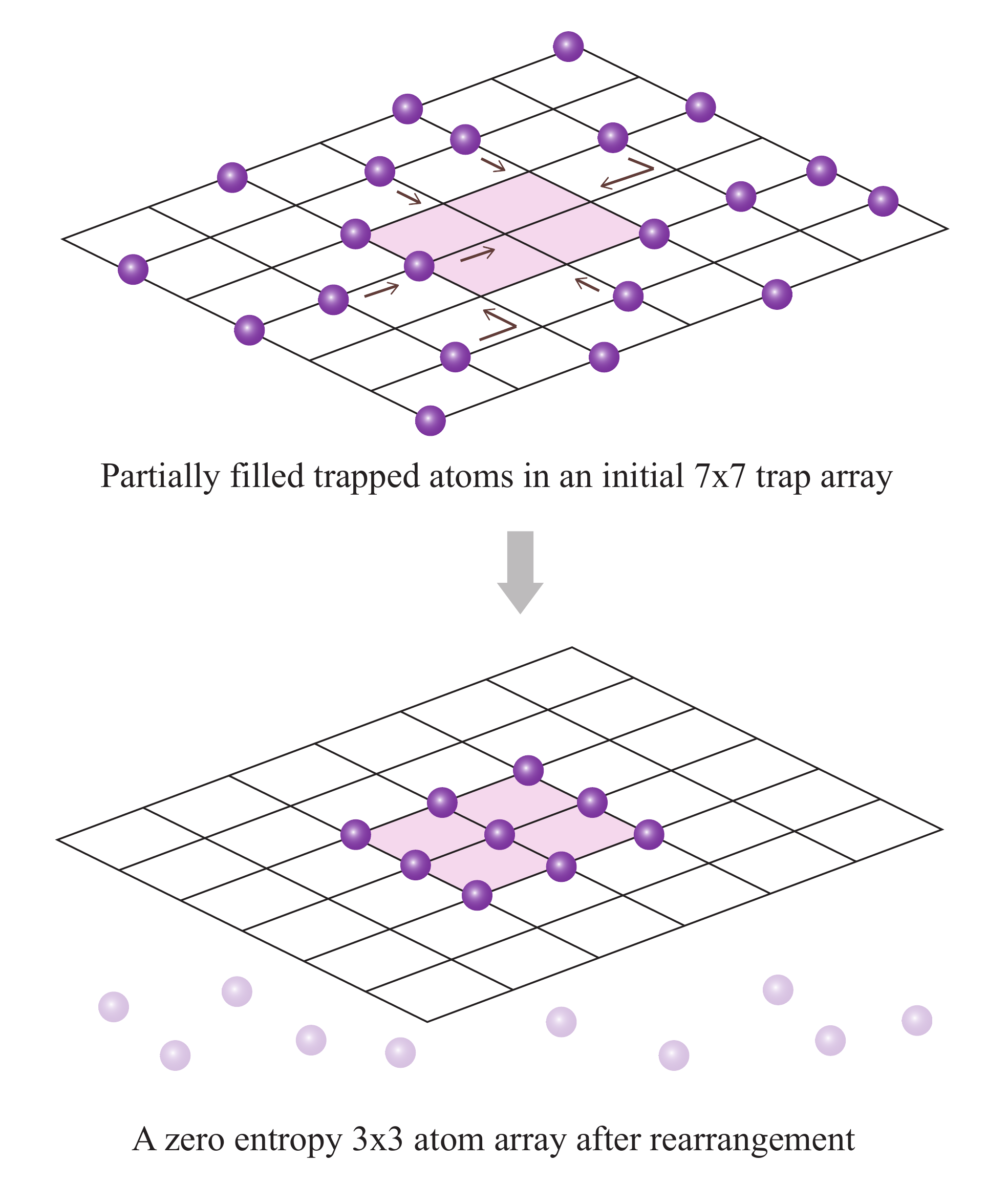}
\caption{Reconfiguration of an initial 7-by-7 lattice of probabilistically-loaded atoms to a completely filled 3-by-3 atomic array. A defect-free array is achieved by filling vacancies with nearby reservoir atoms.}
\label{fig1}
\end{figure}

Recently, methods have been devised to achieve defect-free atom arrays at a high probability by filling vacancies with nearby reservoir atoms~\cite{WeissPRA2004,ValaPRA2005, LeeOE2016, KimNCom2016,  BarredoSci2016, EndresSci2016}, along with the development of atom transport techniques~\cite{KuhrSci2001,MiroshnychenkoNat2006,BeugnonNPhys2007,FatemiOE2007,HeOE2009,LengwenusOE2010}. In this vacancy-filling scheme, as illustrated in Fig.~\ref{fig1}, the probability of achieving $N$ completely filled lattice points is a product of the probability of initially trapping more than or equal to $N$ atoms and the probability of successful transport of $N$ atoms to target sites. Since the former is a conditional probability that approaches one as the number of initial traps exceeds $2N$, the vacancy-filling of the target sites is mainly governed by the latter, or how $N$-atom transport is performed. The shorter the overall travel path of all atoms, the smaller the loss that is given as a function of travel time and distance. Thus, successful transport depends on a  ``good'' atom-guiding plan that minimizes the travel time and distance as well as any lossy transport paths. This is a combinatorial optimization problem, and can be specifically categorized as bipartite matching, for which the solutions can be efficiently found with graph theories such as the Hungarian method, or Hungarian matching algorithm~\cite{Kuhn1955}.

In this paper, we consider the Hungarian matching algorithm as an efficient means to achieve defect-free atomic lattice formation through vacancy-filling. In Sec.~\ref{sec2}, we first compare atom-site matching methods, namely the brute-force and heuristic approaches as well as the Hungarian, to discuss their pros and cons, and then explain how to obtain collision-free paths using the Hungarian algorithm in Sec.~\ref{sec3}. The experimental procedure of capturing atoms with optical-dipole traps, identifying the vacancies, calculating the optimal path plans accordingly, and finally verifying the filling is described in Sec.~\ref{sec4}. In Sec.~\ref{sec5}, we present the results of experiments utilizing the optimal path planning before concluding in Sec.~\ref{sec6}.

\section{Atom-site matching algorithms}
\label{sec2}

When we consider the relocation of atoms to transform a partially-filled atomic lattice to a completely-filled one, finding a set of relocation paths can be viewed as the problem to find a match between every target site and a corresponding atom. Although there are a plethora of algorithms to assign matching between the target sites and the same number of atoms, we must consider their operational efficiency in actual experiments. Not only do atoms in optical-dipole traps have a finite trapping time, but they also escape from the traps during transport with a certain probability given as a function of both time and distance. In choosing a specific algorithm, therefore, we need to consider the time and travel distance. The time is the sum of computational time for the matching algorithm, and execution time for the subsequent guiding operation (transport), with the latter closely related to the travel distance. In our case of about half-filled lattices, the travel distance (or the execution time) does not change much for various initial configurations and algorithms; however, the computation time changes significantly depending on the choice of algorithm. 

\begin{figure}
\centering
\includegraphics[width=0.45\textwidth]{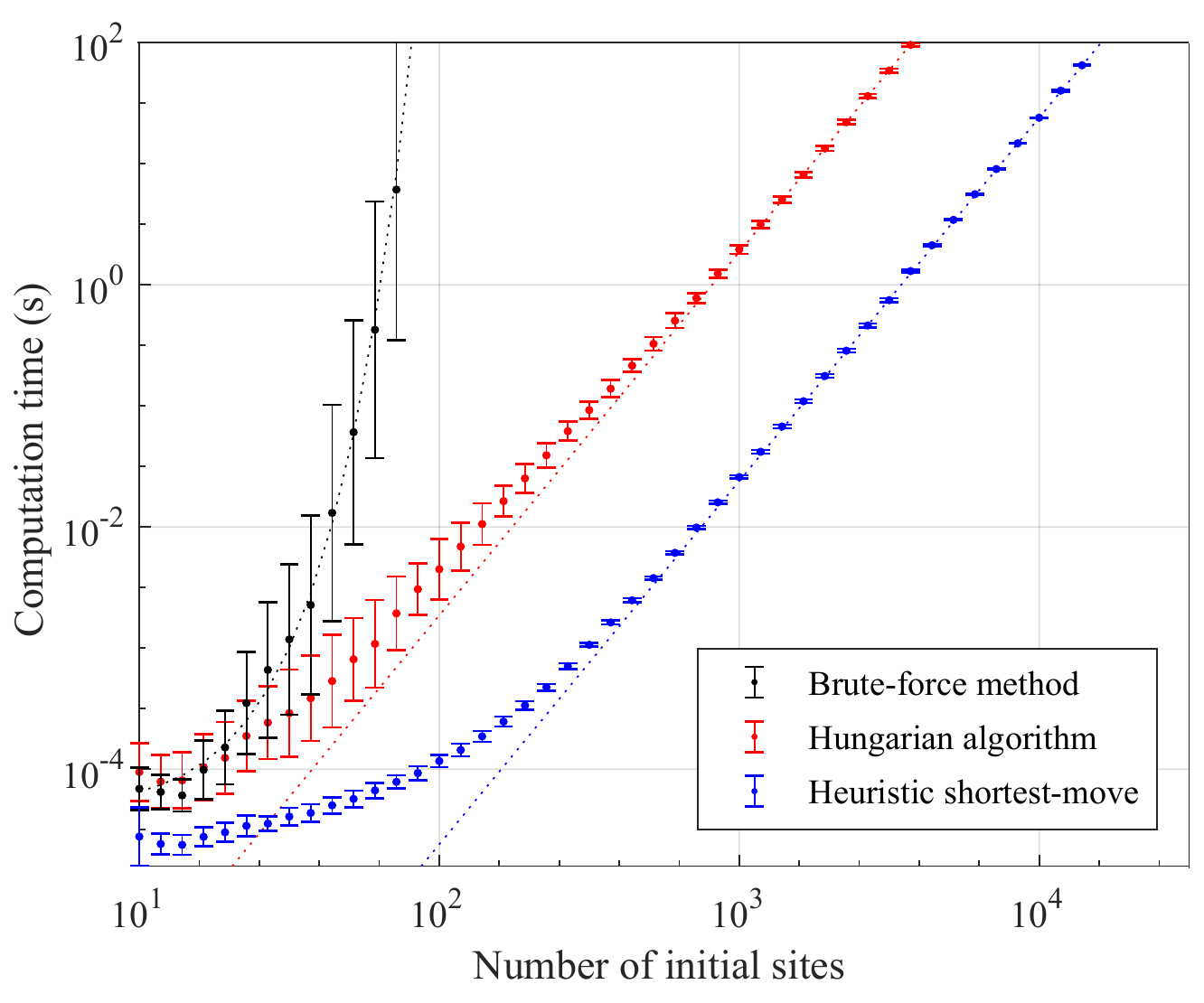}
\caption{Computational time vs. the number of initial sites. The computational time of atom-matching to target sites using the brute-force, heuristic shortest move, and Hungarian algorithms, when the numbers of target sites, atoms and initial sites are given by $N :N_A : N_i=1:2:4$, respectively. Each errorbar represents the standard deviation.}
\label{fig2}
\end{figure}

Figure~\ref{fig2} compares the computational times of various atom-site matching algorithms. The brute-force algorithm requires a factorial increasing computational time as the size of the target site $N$ increases, and the Hungarian algorithm scales as $N^3$~\cite{EdmondsJACM1972}. While the result of the heuristic method (the shortest move method~\cite{BarredoSci2016}) provides a shorter computational time, the resulting matching is not only sub-optimal but also often involves path collisions (see Sec.~\ref{sec3}). The pros and cons of these algorithms are summarized in Table~\ref{table_path}, with the details of each method discussed in the following subsections.
\begin{table}[h]
\caption{%
Comparison of atom-site matching algorithms
}
\label{table_path}
\begin{center}
\begin{tabular}{lccc}
\hline\hline
\textrm{Algorithm}&
\textrm{Calculation complexity}&
\textrm{Rigorosity}\\
\hline
Brute-force method & $O(N!)$ & yes \\
Heuristic shortest-move & $O(N^3)$ & no \\
Hungarian matching & $O(N^3)$ & yes \\
\hline\hline
\end{tabular}
\end{center}
\end{table}

\subsection{Brute-force atom-site matching}  
The brute-force method extensively searches all possible matching solutions; thus, it finds the optimal solution without failure, but in an extremely time-inefficient way. In this method, after identifying the initial configuration of atoms in the lattice, we calculate the distance matrix $D$, of which the element $d_{i,j}$ is the distance between each target site $t_i$ and the initial position of each trapped atom $a_j$. 
When all the target sites are indexed with $T = \{ t_i | 1 \leq i \leq N \}$ and the positions of the trapped atoms with $A = \{ a_j | 1 \leq j \leq N_A \}$, the objective is to find a one-to-one matching $f: T \rightarrow A$ which minimizes the total distance between atoms and target sites, where $d_{\rm total}=\sum_{i}d_{i,f(i)}$ and $d_{i,f(i)}=|t_i-f(t_i)|$. All possible subsets of $A$ of size $N$ are sequentially selected with all possible permutations inspected. This method ensures the optimal solution ({\it i.e.}, the one-to-one function with the minimum total distance); however, it requires a tremendous amount of calculation time. As shown in Fig.~\ref{fig2}, the brute-force calculation time scales factorially as a function of the total number of initial sites, $N_i$, and as a result it takes more than an hour for $N_i=100$, which is not practical in our experiments.

\subsection{Heuristic shortest-move matching}

Heuristic algorithms can find a solution in a time-efficient manner.  One example used in Ref.~\cite{BarredoSci2016}, which may be referred to as heuristic shortest-move matching, finds a solution in such a way that $N$ smallest elements are sequentially selected from the distance matrix $D$ with the condition of choosing only one element from each row and column. So, in the distance matrix, this algorithm finds the smallest element $d_{l,m}$ and assigns $a_m$ to $t_l$, {\it i.e.}, $a_m=f(t_l)$. Then, the $l$'th row and $m$'th column are eliminated from the matrix $D$ and the process repeats $N_T$ times until all target sites are assigned to atoms. This algorithm is fast but not rigorous, with obtained solutions only sub-optimal as it does not give the minimized total distance. As shown in Fig.~\ref{fig2}, the heuristic approach requires a calculation time an order smaller than the Hungarian algorithm for typical cases (30 times faster for $N_i=100$). However, in addition to the drawback of sub-optimal solutions, it also requires additional restriction rules to remove atom-atom collisions en route (see Sec.~\ref{sec3}).
 
\subsection{Hungarian matching algorithm}

Graph theories, such as Hall's marriage theorem, the Hopcroft-Karp algorithm, and the Hungarian algorithm, provide useful theoretical backgrounds to achieve a fast and rigorous matching between target sites and atoms. Hall's marriage theorem~\cite{Hall1935}, or Hall's theorem, provides the necessary and sufficient condition for the existence of a matching $M$ that covers at least one side of a bipartite graph $G(U, V; E)$, where $U$ and $V$ are two finite sets, and $E$ is the set of edges that connect $U$ and $V$. In the current work, we consider $U=T$ and $V=A$, and this theorem tells whether there exists in $G$ any possible exclusive matching between each target site and a corresponding atom among all trapped atoms. The Hopcroft-Karp algorithm~\cite{Hopcroft1973} finds an actual matching $M$ that allows the maximal one-to-one connection between $U$ and $V$, from a given bipartite graph $G$. When all elements in $U=T$ are one-to-one connected to $V=A$, in other words maximal matching, the complete filling of the target sites in our case is possible. This theorem however only finds possible matching, without considering distance minimization.

As total distance minimization is necessary, we focus on the Hungarian matching algorithm, which can use cost functions when finding a maximal matching $M$ in $G$~\cite{Kuhn1955}. The Hungarian method efficiently finds the maximal matching with a time complexity of $N^3$ for an $N \times N$ cost matrix, when the constraint is given to minimize the cost function.  Our Monte Carlo simulation using the total travel distance as the cost function shows the same scaling behavior of computational time as in Fig.~\ref{fig2}. Furthermore, some modifications to the original Hungariam algorithm can significantly reduce the calculation time, either by employing a sparse-matrix Hungarian algorithm or by using the sub-domains of trapped atom sites to apply the algorithm to each domain (a divide-and-conquer approach).

\section{Collision-free path planning by Hungarian algorithm}
\label{sec3}

\begin{figure}[b]
\centering
\includegraphics[width=0.45\textwidth]{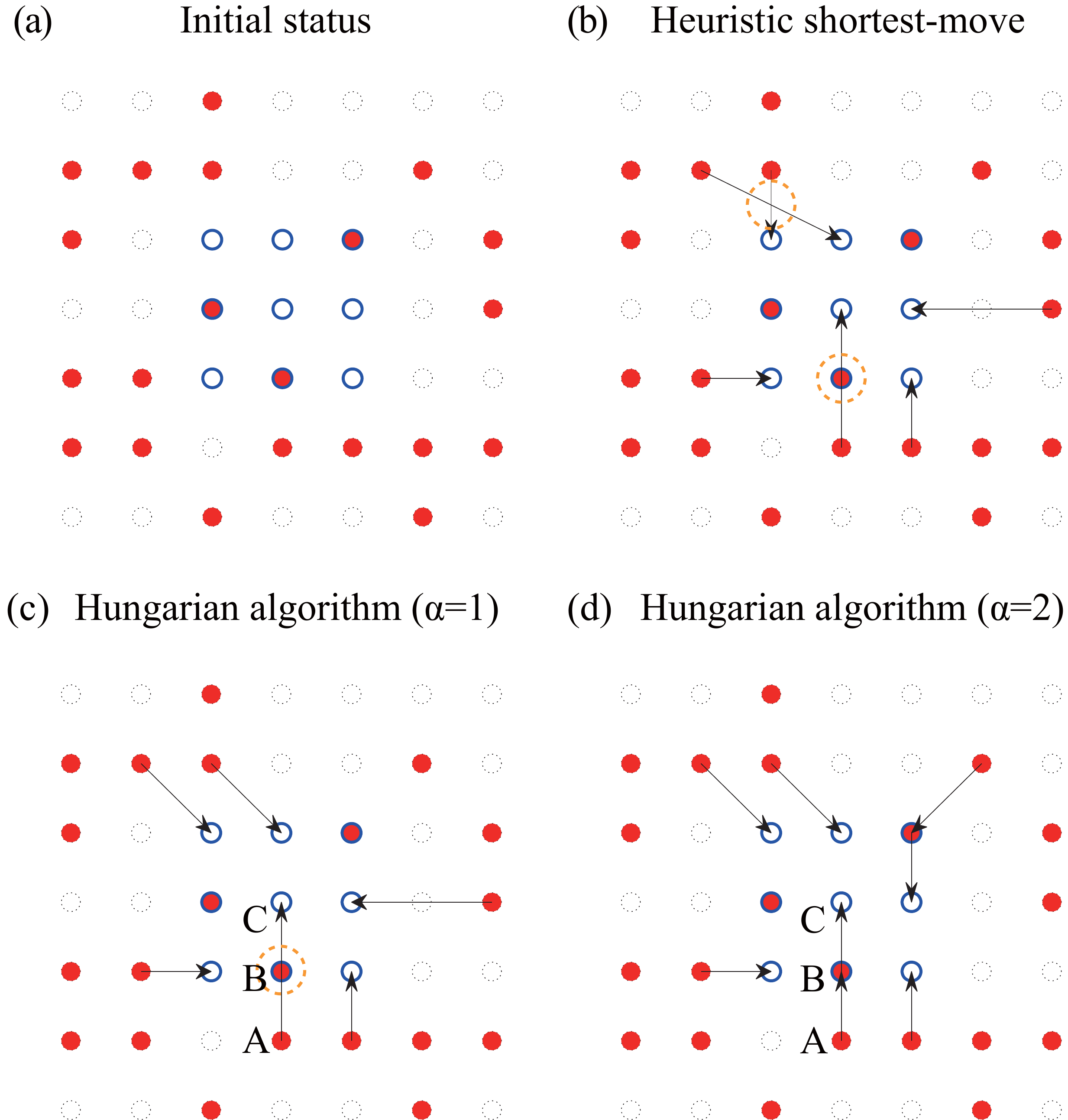}
\caption{Visualization of move solutions from (a) the 7-by-7 initial array, by (b) shortest-move matching algorithm, and (c-d) Hungarian algorithm matching with $\alpha=1$ and $2$, respectively, to the central 3-by-3 target array. The orange dotted circles show overlapping of the paths and trespassing of atom sites.}
\label{fig3}
\end{figure}

Examples of actual atom-guiding plans obtained with the heuristic shortest-move and Hungarian algorithms are shown in Fig.~\ref{fig3}. The initial configuration is a 7-by-7 square lattice ($N_i=49$) randomly occupied by $N_A=21$ atoms, as in Fig.~\ref{fig3}(a), where filled circles represent the initial atoms and unfilled circles the vacancies in the 3-by-3 target lattice ($N=9$). The result of the heuristic shortest-move method is shown in Fig.~\ref{fig3}(b). However, some guiding paths cross each other or trespass on existing atoms (orange dotted circles), which leads to possible atom loss or improper guiding due to the merging of optical-dipole traps en route. 

The Hungarian matching algorithm in Fig.~\ref{fig3}(c), on the other hand, shows no path crossing. This is because the matching with path crossing gives a bigger travel distance than the corresponding collision-free matching that swaps the targets, and the Hungarian algorithm minimizes the total distance. However, trespassing still remains, as shown with the dotted circle in Fig.~\ref{fig3}(c). 
In order to avoid such trespassing, we can employ an alternative cost matrix $D$, for example, with a modified distance metric $d_{i,j}^\alpha$. With the modified distance metric, trespassing is avoided when $\alpha>1$. If, for example $\alpha=2$, since the matching $A \rightarrow B$, $B \rightarrow C$ (``relaying path'') in Fig.~\ref{fig3}(d) gives lower cost ($1^2+1^2=2$) than $A \rightarrow C$, $B \rightarrow B$ (trespass) in Fig.~\ref{fig3}(c) ($2^2+0^2=4$). 
A similar principle can also apply to ``nearly trespassing paths'' where, for instance, atom $B$ is near the $A \rightarrow C$ path. Since atom traps have finite sizes in space, by avoiding the atoms which are too close, atom loss could be reduced. In a similar manner to the trespassing case, a relaying path is chosen when $\alpha>\alpha_c$, in which the minimum interatomic distance is increased. Sufficient $\alpha_c$ can vary according to the array configuration. For the square lattice in our case, it is found that $\alpha > 1.12$ ensures the minimum interatomic distance of $1/\sqrt{2}$, as follows. We consider nearly trespassing configurations that involve the minimum interatomic distance,  in which $(0,0) \rightarrow (1,l)$ and $(0,1) \rightarrow (0,1)$ is the nearly trespassing path.  The condition for the relaying path is $1^\alpha + (\sqrt{1+(l-1)^2})^\alpha < (\sqrt{1+l^2})^\alpha$. For $l=1$, a nearly trespassing path is allowable because the minimum distance in this case is $1/\sqrt{2}$, which is sufficiently larger than the trap size. For $l=2$, 
$\alpha_c\approx1.12$, and as $\alpha_c$ has smaller values for larger $l$'s, $\alpha > \alpha_c$ ensures the minimum distance not to be smaller than $1/\sqrt{2}$, which is the condition for collision-free matching.

\section{Experimental procedure}
\label{sec4}

The experimental setup, similar to what is described in our earlier work~\cite{LeeOE2016, KimNCom2016}, includes a magneto-optical trap (MOT) for cold rubidium atoms ($^{87}$Rb), a dipole-trapping laser beam programmable with a 2D spatial light modulator (SLM, Meadowlarks XY spatial light modulator, 512$\times$512 pixels, 200~Hz frame rate) in the Fourier domain,  a single-atom imaging system with an electron multiplying charge-coupled device (EMCCD) and a high numerical aperture lens (NA = 0.5), and a computing system that calculates possible atom-relocation paths. Atoms were first cooled and trapped in the MOT which took $0.5$ seconds. Simultaneously, the dipole-trapping beams were turned on to prepare an initial array of atoms that were probabilistically loaded in the collisional blockade regime~\cite{SchlosserPRL2002}, with a filling factor of about 50 percent. Then, the imaging system read out the filling and vacancy configuration of the initial atom array, and the computing system calculated an atom-transport path plan to a completely-filled smaller-size lattice. The matching algorithm, such as the Hungarian algorithm, was used at this stage. Once the atom guide plan was finalized, all the atoms to be relocated were simultaneously transported, while the mask pattern for the SLM was calculated in real time, which was accelerated with a graphic processing unit (GPU, Nvidia Titan X). For hologram generation, we used a modified GS (Gerchberg-Saxton) algorithm~\cite{Kim2017}. 
When the first trial of atom reconfiguration was completed, the actual array configuration was confirmed through a second readout. If the configuration was incomplete due to moving or collision loss during the operation, the whole process was repeated until a defect-free array was achieved. The whole experiment was performed in a closed feedback loop with up to nine iterations within the trap lifetime of $\tau=18$~s.

\section{Results and Discussion}
\label{sec5}

\begin{figure}[!t]
\centering\includegraphics[width=0.45\textwidth]{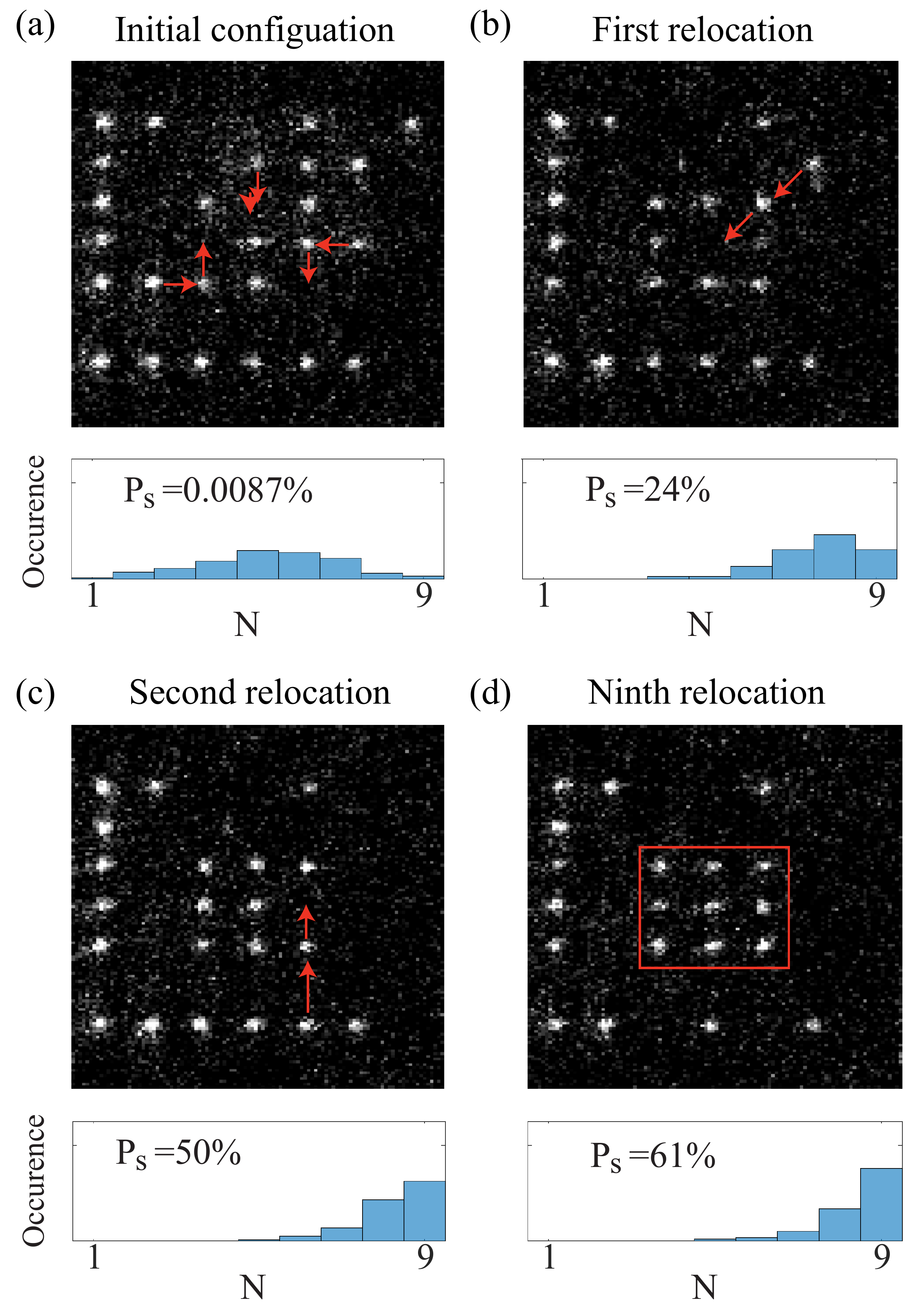}
\caption{Experimental examples of the formation of a 3-by-3 atom array from a partially filled 7-by-7 array using the Hungarian matching algorithm, where $P_s$ is the success probability of achieving a filled target lattice. Histograms of the atom number in the target lattice are shown below from a total of 250 events.}
\label{fig4}
\end{figure}

Experimental demonstration of our defect-free atom-lattice formation using the Hungarian algorithm is shown in Fig.~\ref{fig4}. Representative atom lattice images at various stages are shown in Fig.~\ref{fig4}(a-d). The initial configuration was a partially filled 7-by-7 square lattice, having three vacancies in the central 3-by-3 target zone, as shown in Fig.~\ref{fig4}(a). As indicated by the five arrows, the neighboring atoms were simultaneously moved to construct a completely-filled central lattice. However, as the images in Fig.~\ref{fig4}(b) and (c) show, some atoms in the target lattice disappeared during transport due to time-dependent atom loss. To fill the vacancies, neighboring atoms were additionally moved along the paths indicated with arrows, until a completely filled 3-by-3 lattice was achieved, as shown in Fig.~\ref{fig4}(d). The atom-site matching in each stage and the corresponding guiding paths were obtained using the Hungarian algorithm with $\alpha=1.5$. The success probability $P_s$, defined as the number of successful events (achieving the defect-free 3-by-3 target lattice) divided by the total number of events (250), increased from about $0.5^9$ in the initial configuration, to 24\% after the first relocation, then 50\% after the second relocation, and ultimately 61\% after the ninth relocation.

\begin{figure}
\centering
\includegraphics[width=0.48\textwidth]{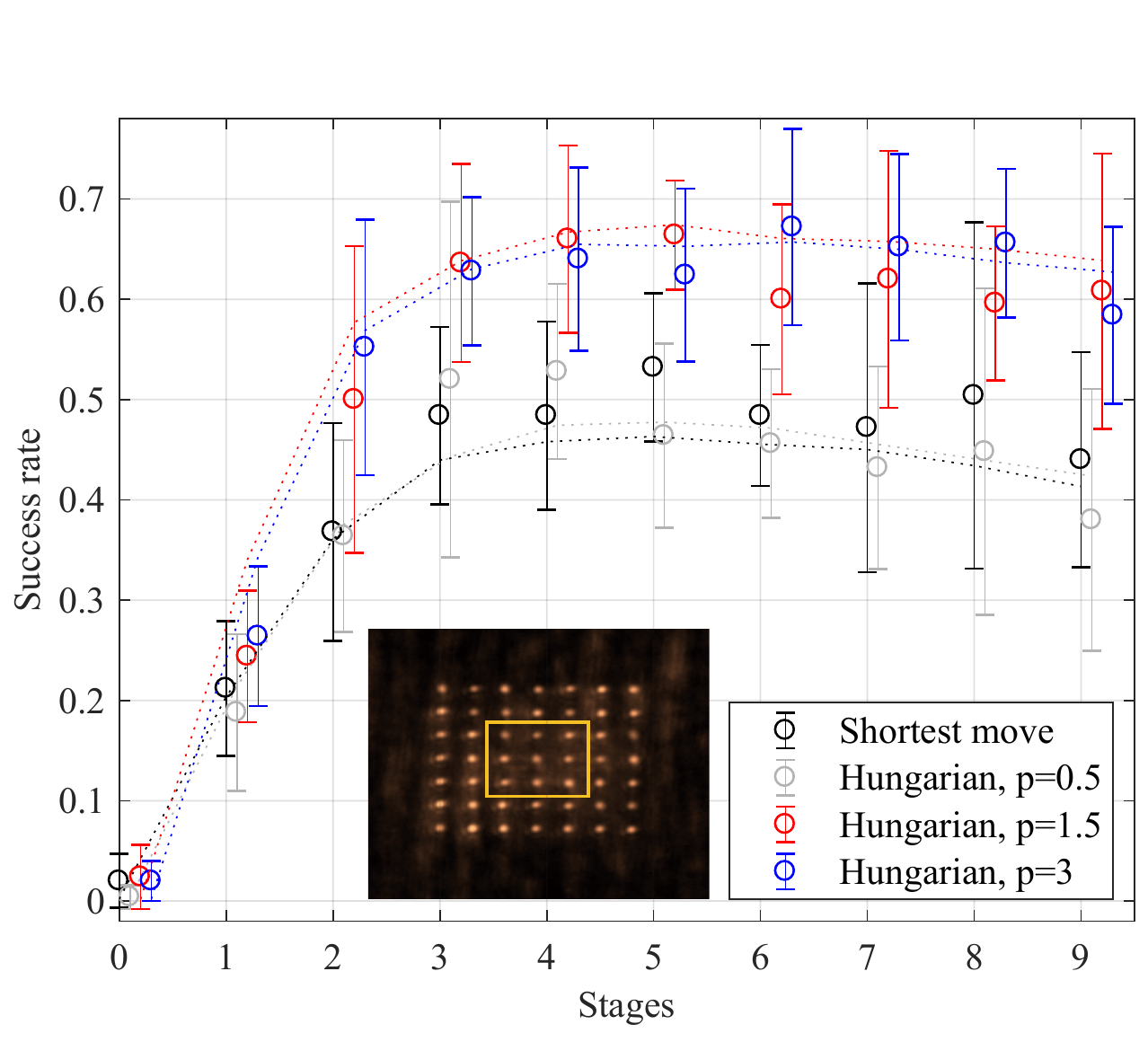}
\caption{Success rate comparison for shortest-move and Hungarian algorithm matching with various \textit{$\alpha$} values in the 7-by-7 lattice for the target 3-by-3 lattice in the central region.}
\label{fig5}
\end{figure}

Figure~\ref{fig5} compares the experimental success probabilities of the Hungarian algorithm  with $\alpha=0,1$, $1.5$ and 3 with that of the heuristic shortest-move matching. The experimental data (circles) shows that the success probability to achieve a defect-free array is notably bigger when the Hungarian matching algorithm with either $\alpha=1.5$ or $\alpha=3$ is employed rather than the heuristic shortest-move algorithm or Hungarian with $\alpha=0.5$. This result is in good agreement with the analysis in Sec.~\ref{sec3}, where it was predicted that the former cases are collision-free but the latter cases are not. Note that the success probability $P_s$ first increases as a function of the stage number, but decreases in the end, which is attributed to the fact that the longer travel distance required for the later stages brings about bigger losses. In the experiment, each atom move between sites was divided into $N_{\rm frame}=15$ segmented moves, and each segmented move was driven by the SLM frame evolution between two stationary frames. The atom survival probability in each segmented move can be modeled as  $P = P_{\rm time} P_{\rm moving} P_{\rm cross}$, where $P_{\rm time}=e^{-t/\tau}$ is the survival probability against the time-dependent loss due to background gas collision, with $\tau$ the trap lifetime,  $P_{\rm moving}=e^{- \beta N_{\rm frame} d^2}$ is the survival probability against the moving loss due to intensity flickering of the optical dipole traps, with $\beta$ the moving loss coefficient and $d$ the travel distance, and $P_{\rm cross}=1-e^{-\gamma d_{\rm min}^2}$ is the survival probability against the loss due to path collisions. In Fig.~\ref{fig5}, the numerical simulation using the above models (dotted lines) for each relocation stage are shown. The fitted parameters obtained through curve fitting are given by $\tau=18$~s, $\beta=0.0076/a^2$, and $\gamma=80/a^2$, where $a$ is the lattice constant. Each data point is statistically averaged over 250 events, where the errorbar represents the standard deviation.

\begin{figure*}
\centering
\includegraphics[width=0.95\textwidth]{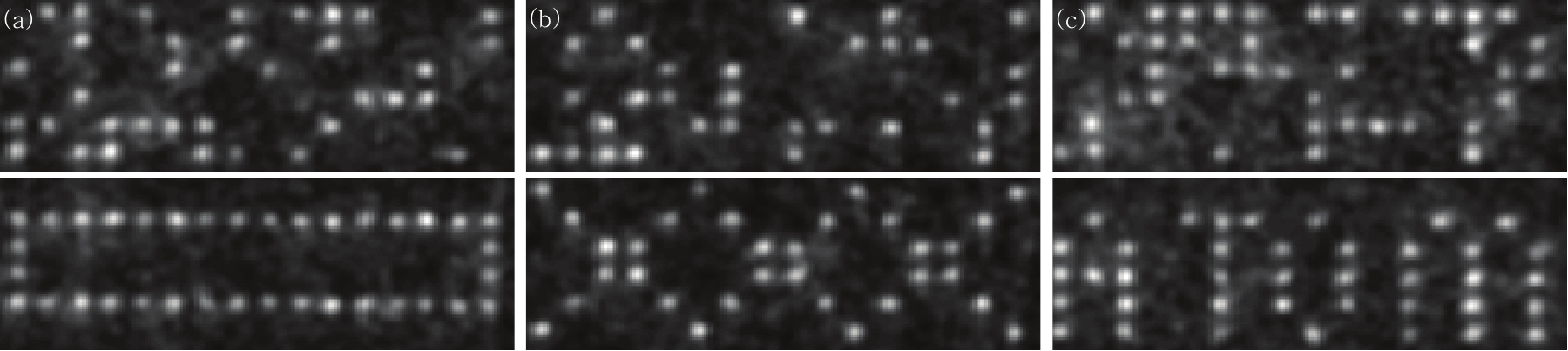}
\caption{Examples of defect-free atomic array formation: (a) a rectangular ring, (b) a triple X, and (c) the capital letters of the word ``atom'' with upper and lower images showing the initial and final configurations, respectively. }
\label{fig6}
\end{figure*}

Finally, Fig.~\ref{fig6} shows a few examples of atom arrays formed by the Hungarian matching algorithm. In each of the demonstrations, the upper images show examples of the random initial configurations with $N\sim100$ initial sites, and the lower images show the final configurations following atom relocation. Each image was from a single shot. For array-formation of various geometries of target sites, the Hungarian algorithm performed well.

\section{Conclusion}
\label{sec6}
Three methods of vacancy site filling have been compared. The advantage of Hungarian matching over the brute-force method is clear because the calculation time of the latter greatly exceeds the former as the number of atoms increases. The heuristic shortest-move method seemingly has an advantage in short calculation time, but the issue of path collisions becomes serious, in particular when the vacancy occurs in the central region of the target lattice. It is concluded that the Hungarian matching method has at least three advantages over the heuristic shortest-move method: it provides rigorous solutions, high success probabilities, and advantages in atom vacancy healing cases, where the second and third advantages are attributed to the collision-free path planning of the Hungarian algorithm.

\begin{acknowledgements}
This research was supported by Samsung Science and
Technology Foundation [SSTF-BA1301-12]. 
\end{acknowledgements}
\FloatBarrier

\end{document}